# Steps towards nonlinear cluster inversion through gravitational distortions.
# II. Generalization of the Kaiser & Squires method

Carolin Seitz & Peter Schneider

## Abstract

The weak distortions of high-redshift galaxies caused by gravitational light deflection near clusters of galaxies can be used to reconstruct the projected (two-dimensional) surface mass density of intermediate redshift clusters. This technique, pioneered by Tyson, Valdes & Wenk, and Kaiser & Squires, is reconsidered in the present paper, where we generalize the inversion equation found by Kaiser & Squires (KS) in several respect. Adopting a different smoothing procedure for the discreetly sampled data (individual galaxy images), we effectively reduce the shot noise in the KS procedure. In particular we show that the best density reconstructions are obtained if the smoothing scale is adopted to the 'strength of the signal', which yields a better resolution near the center of the cluster where the distortions are strongest. Furthermore, we point out the importance of boundary effects and demonstrate their disastrous impact on rectangular data fields (CCDs) with large side ratio. Most important, however, is the generalization of the KS method to critical clusters, i.e., to such clusters which are capable of producing multiple images and giant luminous arcs. The corresponding modifications of the inversion procedure are severe; in particular, the resulting inversion equation is much more difficult to solve. As we pointed out in a previous paper (Schneider & Seitz), there exists a local degeneracy if the cluster is critical. We have developed an iteration procedure to solve the inversion equation, which we demonstrate to yield a very accurate reconstruction of the cluster mass distribution. In particular, the mass within the inner few arcminutes of the cluster can be determined with an error of only a few percent, thus demonstrating the potential applicability and accuracy of this method for cluster mass determinations.

# 1. Introduction

The investigation of the distortion of faint background galaxies due to the tidal gravitational light deflection has become one of the most promising methods for the determination of the mass distribution in clusters of galaxies. The strong distortions manifested in giant luminous arcs (see, e.g., the review by Fort & Mellier 1994), and the weakly distorted, but much more numerous images of background galaxies (Tyson, Waldes & Wenk 1990) provide a unique tool for putting strong constraints on the mass profile in clusters, which are not based on simplifying assumptions like statistical equilibrium of the cluster galaxies, or the degree of anisotropy of the orbits of the galaxies. Methods for the determination of the radial profile of (nearly) spherical clusters have been investigated by Kochanek (1990), Miralda-Escudé (1991) and others. For the cluster Abell 370, the giant luminous arc, several arclets, and several candidate image pairs have yielded a very detailed mass model (Kneib et al. 1993). The consideration of weak distortions in the outer parts of the cluster CL0024+1654 has yield to the determination of the mass profile in the outer parts of this cluster (Bonnet, Mellier & Fort 1994). In their pioneering paper, Kaiser & Squires (1993, hereafter KS) have found an inversion formula for the mass profile of a cluster in terms of the shear distribution, which can be obtained from the distortion of the images of background sources. This reconstruction technique has been applied to several clusters (Fahlman et al. 1994, Smail et al. 1994,b); one of the exciting results was that the so-determined mass in the cluster ms1224 (Fahlman et al. 1994) is about three times larger than that predicted if mass traces light, with a mass-to-light ratio determined from a virial analysis.

In a previous paper (Schneider & Seitz 1994, hereafter Paper I) we have investigated the image distortions from gravitational lensing, dropping the assumption of weak lensing, which is implicitly assumed in the KS method. There we showed that the shear is not an observable quantity, but that the only local observable from image distortion is the distortion $\delta$ (defined in Sect. 2 below). We have investigated the statistical properties of the distortion and derived several useful statistical estimates of $\delta$ from a given set of distorted images. We have applied our results to spherical clusters and obtained very accurate density reconstructions, even for the case that the cluster is critical, i.e., capable of producing multiple images.

In this paper we generalize our method to the case of a realistic two-dimensional mass distribution. As in Paper I we confine our considerations to the case that all sources are at the same redshift, to concentrate on the essentials of our new method. In Sect. 2 we introduce our notation and derive an inversion formula for the surface mass density in terms of the distribution of the distortion, which for the case of weak lensing reduces to the inversion equation of KS. The discreteness of galaxy images implies that the density profile of the cluster can be obtained only with a finite resolution, since a smoothing procedure has to be applied. In Sect. 3 we consider the case of weak lensing and compare the smoothing procedure proposed by KS with a new one. It will be demonstrated that our new smoothing procedure yields considerably better results compared to that used by KS, since it essentially removes the shot-noise contribution from the local error. In particular, we show that the smoothing scale can be adapted to the signal strength to yield higher resolution in those regions of the cluster where the distortion is stronger, i.e., in the central part of the cluster. For our numerical examples, we use a model cluster as obtained from cosmological $N$-body simulations for a CDM universe, as described in Bartelmann, Steinmetz & Weiss (1994); this cluster is part of a sample of model clusters used by these autors to investigate the statistical properties of arcs. We also investigate some aspects of the boundary effects which arise from the finite field of observation, and show that some of the features seen in the reconstructed profile of the cluster 0016+16 (Smail et al. 1994) are just artifacts of using data from a rectangular field. In Sect. 4 we use the full inversion equation to reconstruct the mass profile of the cluster. In contrast to the weak lensing case, where the density as a function of position can be calculated from an integral over the (weak) distortion, multiplied by a simple kernel, this case now involves the solution of an integral equation on a finite region of the (lens) plane. As was already discussed in depth in Paper I, there is a local degeneracy in the estimate of the kernel if the cluster is critical. This local degeneracy cannot be broken by local observations, but can only be removed by applying smoothness conditions on the shear distribution. We discuss the resulting difficulties in detail and provide a reconstruction of the cluster profile from synthetic data of galaxy image distortions. It will be demonstrated that this method yields fairly accurate density profiles even for clusters with supercritical surface mass density, and that even some weak substructure is reobtained from the reconstruction. Finally, we discuss our results in Sect. 5.

## 2. The generalized Kaiser & Squires method

### 2.1 The inversion equation

As already pointed out, we make the simplifying assumption that all sources have the same redshift; hence, the critical surface mass density

$$\Sigma_{\rm crit} = \frac{c^2 D_{\rm s}}{4\pi G D_{\rm d} D_{\rm ds}} \ . \tag{2.1}$$

is the same for all sources, where $D_{\rm d}$ and $D_{\rm s}$ are the angular diameter-distances from the observer to the lens and to the sources, and $D_{\rm ds}$ is the angular diameter-distance from the lens to the sources.[1] The dimensionless surface mass density is then defined as

$$\kappa(\boldsymbol{\theta}) = \frac{\Sigma(D_{\rm d}\boldsymbol{\theta})}{\Sigma_{\rm crit}}, \tag{2.2}$$

where $\boldsymbol{\theta}$ denotes the angular position of the light ray in the lens plane, and $\Sigma(D_{\rm d}\boldsymbol{\theta})$ is the physical surface mass density of the lens.

The deflection potential

$$\psi(\boldsymbol{\theta}) = \frac{1}{\pi} \int_{\mathbb{R}^2} {\rm d}^2\theta' \ \kappa(\boldsymbol{\theta}') \ \ln\left|\boldsymbol{\theta} - \boldsymbol{\theta}'\right| . \tag{2.3}$$

---

[1]   More precisely, we assume that the ratio $D_{\rm ds}/D_{\rm s}$ is the same for all sources; this assumption is supposed to be not too bad for low-redshift clusters (e.g., $z_{\rm d} \sim 0.2$), where this ratio is very insensitive to the source redshifts, provided they are larger than about three times the cluster redshift, or $z_{\rm s} \gtrsim 0.6$ in our example. If the galaxy sample is sufficiently faint, which it must be anyway to yield a sufficiently large number density of sources, this assumption on the source redshifts will be satisfied reasonably well.

satisfies the two-dimensional Poisson's equation (for notation and further discussion, see Chap. 5 of Schneider, Ehlers & Falco 1992)

$$\kappa(\boldsymbol{\theta}) = \frac{1}{2}\nabla^2\psi(\boldsymbol{\theta}) = \frac{1}{2}\left(\psi_{11}(\boldsymbol{\theta}) + \psi_{22}(\boldsymbol{\theta})\right) , \tag{2.4}$$

with $\psi_{ij}(\boldsymbol{\theta}) = \frac{\partial^2}{\partial\theta_i \partial\theta_j}\psi(\boldsymbol{\theta})$.

The lens equation

$$\boldsymbol{\beta} = \boldsymbol{\theta} - \nabla\psi(\boldsymbol{\theta}) , \tag{2.5}$$

describes the angular position $\boldsymbol{\beta}$ of the light ray in the source plane as a function of its angular position $\boldsymbol{\theta}$ in the lens plane, and the distortion of images is described by the Jacobian matrix,

$$A(\boldsymbol{\theta}) \equiv \frac{\partial\boldsymbol{\beta}}{\partial\boldsymbol{\theta}} = \begin{pmatrix} 1 - \kappa - \gamma_1 & -\gamma_2 \\ -\gamma_2 & 1 - \kappa + \gamma_1 \end{pmatrix} , \tag{2.6}$$

with components

$$\begin{aligned}\gamma_1 &= \frac{1}{2}(\psi_{11} - \psi_{22}) , \\ \gamma_2 &= \psi_{12} = \psi_{21}\end{aligned} \tag{2.7}$$

Combining the two components of the shear into a complex number,

$$\gamma = \gamma_1 + \mathrm{i}\gamma_2 , \tag{2.8}$$

and inserting (2.3) into (2.7), we obtain for the complex shear

$$\gamma(\boldsymbol{\theta}) = -\frac{1}{\pi}\int_{\mathbb{R}^2} \mathrm{d}^2\theta'\, \mathcal{D}(\boldsymbol{\theta} - \boldsymbol{\theta}')\, \kappa(\boldsymbol{\theta}') , \tag{2.9}$$

where

$$\mathcal{D}(\boldsymbol{\theta}) = \frac{\theta_1^2 - \theta_2^2 + 2\mathrm{i}\theta_1\theta_2}{|\boldsymbol{\theta}|^4} \tag{2.10}$$

is the complex kernel in (2.9). The inversion of (2.9) reads

$$\kappa(\boldsymbol{\theta}) = \frac{-1}{\pi}\int_{\mathbb{R}^2} \mathrm{d}^2\theta'\, \mathcal{R}\mathrm{e}\left[\mathcal{D}(\boldsymbol{\theta} - \boldsymbol{\theta}')\, \gamma^*(\boldsymbol{\theta}')\right] , \tag{2.11}$$

where $\mathcal{R}\mathrm{e}(x)$ means the real part of the complex number $x$, and the asterisk denotes complex conjugation. The convolution-type integral in (2.9) suggests that this inversion can be easily performed in Fourier space, as was done by KS, but can also be verified directly by insertion. We should note that this inversion is not unique but adding a homogeneous sheet of matter to $\kappa(\boldsymbol{\theta})$ does not change the shear $\gamma$ in (2.9).

Eq. (2.11) shows that we could reconstruct the surface mass density $\kappa(\boldsymbol{\theta})$ of the lens, if the shear $\gamma(\boldsymbol{\theta})$ caused by the deflector can be measured locally as a function of angular position $\boldsymbol{\theta}$. Unfortunately, however, the shear is not an observable quantity in general, as was pointed out in Paper I. There we showed that the only quantity that can be deduced from local observations of image distortions (see also further below) is the complex distortion

$$\delta = \frac{2\gamma(1-\kappa)}{(1-\kappa)^2 + |\gamma|^2} . \tag{2.12}$$

Thus, only this combination of the surface mass density and the shear can be obtained. Solving for the shear

$$\gamma = \frac{(1-\kappa)}{\delta^*}\left[1 \pm \sqrt{1-|\delta|^2}\right] \quad , \tag{2.13}$$

we see that for a given value of $\delta$ and $\kappa$ two solutions of $\gamma$ exist. The (local) degeneracy caused by the two signs in (2.13) cannot be resolved from a local measurement of image distortions. The correct sign is the negative of the sign of $\det A(\boldsymbol{\theta})$; however, the sign of $\det A$, or the parity of local lens mapping, cannot be observed locally. Inserting (2.13) into (2.11) yields

$$\kappa(\boldsymbol{\theta}) = \frac{-1}{\pi}\int_{\mathbb{R}^2} \mathrm{d}^2\theta' \, \mathcal{R}\mathrm{e}\left[\mathcal{D}(\boldsymbol{\theta}-\boldsymbol{\theta}')\left(\frac{1\pm\sqrt{1-|\delta(\boldsymbol{\theta}')|^2}}{\delta(\boldsymbol{\theta}')}\right)(1-\kappa(\boldsymbol{\theta}'))\right] \quad . \tag{2.14}$$

With this integral equation we can reconstruct the surface mass density $\kappa(\boldsymbol{\theta})$, provided that we can obtain $\delta(\boldsymbol{\theta})$ from observations, and that we make the correct choice of the sign in (2.14). As we shall see below, this latter point is one of the main theoretical difficulties for the inversion of clusters which have critical curves (i.e., which are capable to produce multiple images). For non-critical clusters ($\det A > 0$ everywhere), the negative sign has to be taken.

### 2.2 Local observables and their measurement

Next we briefly summarize how the distortion $\delta$ can be obtained from observations of extended galaxy images mapped throught the deflector; for details see Paper I. The shape of a source is described by the complex ellipticity

$$\chi^{(s)} = \frac{(Q^{(s)}_{11} - Q^{(s)}_{22}) + 2\mathrm{i}Q^{(s)}_{12}}{Q^{(s)}_{11} + Q^{(s)}_{22}} \quad , \tag{2.15}$$

where $Q^{(s)}_{ij}$ (Eq.(2.2) in Paper I) are the components of the tensor of second moments of the surface brightness distribution of the source. For source with elliptical isophotes and axis ratio $r \leq 1$, the absolute value of the complex ellipticity $\chi^{(s)}$ is $(1-r^2)/(1+r^2)$, and the phase of $\chi^{(s)}$ yields the orientation of the source. Defining in analogy to that the complex ellipticity $\chi$ of the image, we found for the transformation between $\chi^{(s)}$ and $\chi$

$$\chi^{(s)} = \frac{2g + \chi + g^2\chi^*}{1 + |g|^2 + 2\mathcal{R}\mathrm{e}(g\chi^*)} \quad , \tag{2.16}$$

where

$$g = \frac{\gamma}{(1-\kappa)} \quad . \tag{2.17}$$

Some aspects of the determination of the image ellipticities from observations have been investigated in depth by Bonnet & Mellier (1994). The basic assumption underlying all cluster inversion techniques is that the sources are intrinsically oriented randomly, i.e.

$$\left\langle \chi^{(s)} \right\rangle = 0 \quad , \tag{2.18}$$

where the average is taken over an ensemble of sources. For a finite number of sources, the average (2.18) will not be exactly zero, unless the intrinsic source ellipticity vanishes. The discreteness of sources implies that for a local estimate of lens parameters, one has to average over several galaxy images. Hence, for a given position $\boldsymbol{\theta}$ in the cluster, we consider all galaxy images in a circle of angular radius $\Delta\theta$, and apply (2.18) to them, i.e.

$$\sum_{i=1}^{N_{\mathrm{g}}} \frac{2g + \chi_i + g^2 \chi_i^*}{1 + |g|^2 + 2\mathcal{R}\mathrm{e}(g\chi_i^*)} = 0 \quad . \tag{2.19}$$

A solution of (2.19) yields a local estimate for the lens parameter $g(\boldsymbol{\theta})$. However, as was discussed in Paper I, if $g$ is a solution of (2.19), so is $1/g^*$; this is the basic reason for the occurrence of the local (sign) degeneracy in (2.14). Hence, the only parameter that can be determined locally is the distortion $\delta$, which in terms of $g$ reads $\delta = 2g/(1+|g|^2)$. Then, Eq. (2.19) can be simplified to

$$\sum_{i=1}^{N_{\mathrm{g}}} \frac{\delta + \chi_i}{1 + \mathcal{R}\mathrm{e}(\delta \chi_i^*)} = 0 \ , \tag{2.20}$$

which can be solved iteratively, starting from the 'linear' approximation $\delta \approx -\langle\chi\rangle$. Thus, one can derive a statistical estimate for $\delta$ at any location in the cluster using the ellipticities of the few nearest images, which leads to a smoothed distribution of the distortion $\delta$. The rms error in determining the distortion is $\sigma(\delta) \approx \sigma_\chi(\delta)/\sqrt{N_{\mathrm{g}}}$ with $\sigma_\chi$ shown in Fig. 2 of Paper I; see also Fig. 5 of Paper I. In practice, we performed the smoothing in (2.20) by applying Gaussian weights, i.e.,

$$\sum_{i=1}^{N_{\mathrm{g}}} \exp\left(-\frac{(\boldsymbol{\theta} - \boldsymbol{\theta}_i)^2}{(\Delta\theta)^2}\right) \frac{\delta(\boldsymbol{\theta}) + \chi_i}{1 + \mathcal{R}\mathrm{e}(\delta(\boldsymbol{\theta})\chi_i^*)} = 0 \ . \tag{2.21}$$

We would like to point out that $\Delta\theta$ need not be a constant, but can be chosen to depend on $\boldsymbol{\theta}$ – see below.

### 2.3 The model cluster and its distortion field $\delta(\boldsymbol{\theta})$

Throughout this paper we use one model cluster obtained from cosmological N-body simulations for a CDM universe, as described in Bartelmann, Steinmetz & Weiss (1994) and used for studies of arc statistics by Bartelmann & Weiss (1994) and Bartelmann, Steinmetz & Weiss (1994). For our purposes, we use only the shape of the projected, i.e., two-dimensional mass distribution, and have chosen the scale (i.e., the critical surface mass density) such that the cluster has a central density of $\kappa_0 \approx 1.5$, so that it is critical. The field of the cluster that we consider is a square of about 10'. Fig. 1 shows this dimensionless surface mass density $\kappa(\boldsymbol{\theta})$ for coordinates in units of arcminutes. Besides the most massive central structure, the cluster clearly shows another three less massive substructures.

Then we use the surface mass density $\kappa(\boldsymbol{\theta})$ and the corresponding shear $\gamma(\boldsymbol{\theta})$, calculated directly from the mass distribution of the $N$-body simulations via (2.9), to generate a distribution of synthetic galaxy images. For that, we distribute $N_{\mathrm{gal}} = 3360$ galaxies (sources), i.e., corresponding to a density of about $n = 35\,\mathrm{galaxies/arcminutes}^2$, with random positions and random ellipticities $\chi^{(s)}$ drawn from the probability density

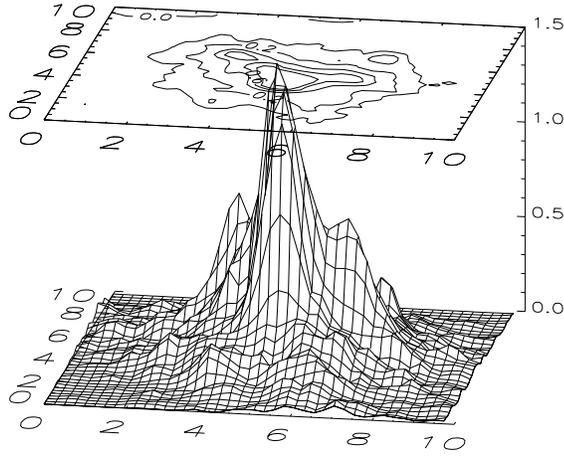

**Fig. 1.** The projected dimensionless surface mass density $\kappa$ of the numerically modelled cluster. The sidelength of the cluster is about 10'. The levels of the contourlines are 0.1,0.2,0.4,0.6,0.8

$$p_s\left(\left|\chi^{(s)}\right|\right) = \frac{1}{\pi R^2 \left(1 - e^{1/R^2}\right)} e^{-\left|\chi^{(s)}\right|^2/R^2}, \quad \text{with} \quad R = 0.3, \qquad (2.22)$$

and calculate the 'observable' image ellipticity $\chi$ for each galaxy from the inverse of Eq. (2.16). In Eq. (2.22) we defined the probability density $p_s\left(\chi^{(s)}\right)$ such that $p_s\left(\chi^{(s)}\right) d^2\chi^{(s)}$ is the probability that the source has an ellipticity within $d^2\chi^{(s)}$ of $\chi^{(s)}$.

Then, using the 'observed' galaxy ellipticities $\chi_i$ we deduce the distortion $\delta(\boldsymbol{\theta})$ on a quadratic grid with $N \times N$ gridpoints $\boldsymbol{\theta}_{ij}$ (with $N = 41$, so that the gridpoint separation is about 0'.25). For each column $j$, we start with the angular smoothing radius $\Delta\theta \approx 0'.3$ (see Eq. (2.21); all lengthscales are given in arcminutes) at the boundary, calculate $\delta$ and use

$$\Delta\theta_{ij} = \frac{\left(1.05 - |\delta(\boldsymbol{\theta}_{i,j-1})|^2\right)^{0.75}}{\sqrt{40}} 1.8, \qquad (2.23)$$

for the following gridponts of the column, where $\delta(\boldsymbol{\theta}_{i,j-1})$ is the distortion found at the previous gridpoint. With this smoothing length the absolute rms error $\sigma_\chi(\delta)$ (see Eq.(2.23a) in Paper I) in determing $\delta$ is approximately the same for all $\delta$, since $\sigma_\chi(\delta) \propto (1-\delta^2)^{0.75}/\sqrt{N_g}$ and $N_g \propto n(\Delta\theta_{ij})^2$ ($n \approx 35$ in our simulations). Thus, we average over many images ($N_g \approx 10$) in regions with small distortion ($|\delta| \lesssim 0.1$) and large dispersion $\sigma_\chi(\delta)$ (Fig. 2 of Paper I), whereas in regions of large distortions ($|\delta| \approx 1$) and small dispersion $\sigma_\chi(\delta)$ we average over very few galaxy images. The procedure described above gives a fairly good reconstruction of the distortion $\delta$, as can be seen by comparing Figs. 2a&b, where we show contour-lines for the original and reconstructed distortion-component $\delta_2 = \mathcal{I}\mathrm{m}(\delta)$. We find that the noise is larger in regions of smaller distortions, because with the chosen smoothing length (2.23) the relative error in determining $\delta$ is larger for regions of smaller distortions. Therefore instead of (2.23) one could also use a smoothing function which provides approximately the same relative error in the reconstruction of $\delta$. For the smoothing function (2.23) we used information about the rms error $\sigma_\chi$ which we derived in PaperI for source ellipticity distributions according to Eq. (2.22). But we point out that the exact form of the smoothing length is not important, but it has to be adopted such that the smoothing is small in regions of high distortions and large in regions of small distortions. That means that we need no information about the ellipticity distribution of the sources to obtain the distortion $\delta$ for the data field.

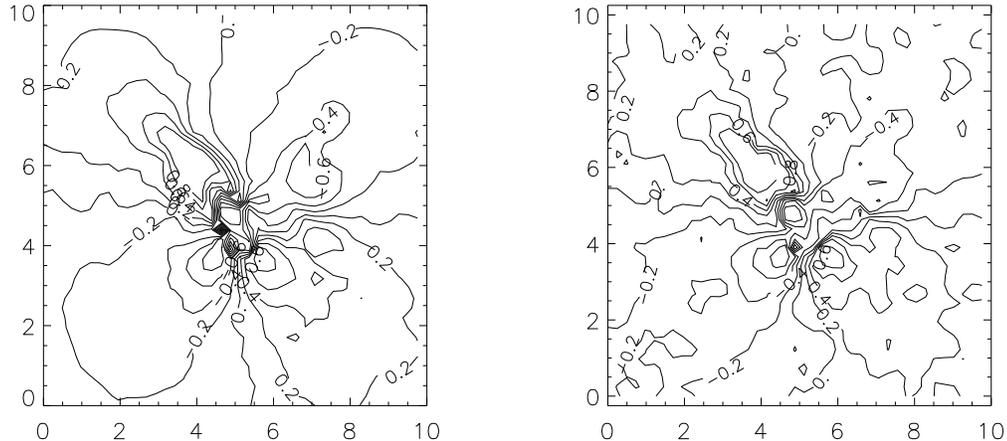

**Fig. 2.** Two contour plots of the distortion-component $\delta_2 = \mathcal{I}m(\delta) = 2\gamma_2(1-\kappa)/[(1-\kappa)^2 + |\gamma|^2]$. The levels of the contour lines are $-0.8, -0.6, \ldots, 0.8$. (a) (left panel): $\delta_2$ calculated from the surface mass distribution of the model cluster (Fig. 1). (b) (right panel): $\delta_2$ calculated from the synthetic galaxy images using Eqs.(2.21) & (2.23)

## 3. The weak lensing regime and the Kaiser & Squires inversion technique

In this section we apply Eq.(2.11) to the weak lensing regime. In this case, the resulting inversion procedure is similar to that derived by KS. We show that their inversion formula can be improved considerably with little effort.

In the weak lensing regime, $\kappa \ll 1$, $\gamma \ll 1$, the distortion in Eq.(2.12) reduces to

$$\delta \approx 2\gamma \ . \qquad (3.1)$$

The relation between $\gamma$ and $\delta$ is unambigous now and the surface mass density $\kappa(\boldsymbol{\theta})$ is determined by the integral

$$\kappa(\boldsymbol{\theta}) \approx \frac{-1}{2\pi} \int_{\mathbb{R}^2} \mathrm{d}^2\theta' \ \mathcal{R}e\left[\mathcal{D}(\boldsymbol{\theta} - \boldsymbol{\theta}') \delta^*(\boldsymbol{\theta}')\right] \quad . \qquad (3.2)$$

Since the distortion is related to the expectation value of the image ellipticity by

$$\delta = -\frac{\langle \chi \rangle}{\zeta(|\delta|)} \ , \qquad (3.3)$$

(see Eq.(3.16) and Fig. 1 of Paper I), and since $\zeta$ is between $0.95 \lesssim \zeta(|\delta|) \leq 1$ for an ellipticity distribution of the form (2.22) with $R = 0.3$, one can use the approximation

$$\delta \approx -\langle \chi \rangle \ . \tag{3.4}$$

This approximation is fairly accurate (see Fig. 1 in Paper I) for intrinsic ellipticity distributions $p_s\left(|\chi^{(s)}|\right)$ with low dispersion. The resulting equation for $\kappa$ reads

$$\kappa(\boldsymbol{\theta}) \approx \frac{1}{2\pi} \int_{\mathbb{R}^2} \mathrm{d}^2\theta' \, \mathcal{R}\mathrm{e}\left[\mathcal{D}(\boldsymbol{\theta} - \boldsymbol{\theta}') \, \langle\chi\rangle^*(\boldsymbol{\theta}')\right] \ . \tag{3.5}$$

Unfortunately, one can not measure $\langle\chi\rangle(\boldsymbol{\theta})$ accurately, due to the finite density of galaxy images. Because of this, KS converted (3.5) to a sum over galaxies

$$\kappa(\boldsymbol{\theta}) \approx \frac{1}{2\pi n} \sum_{k=1}^{\mathrm{Ngal}} W(\boldsymbol{\theta} - \boldsymbol{\theta}_k; s) \, \mathcal{R}\mathrm{e}\left[\mathcal{D}(\boldsymbol{\theta} - \boldsymbol{\theta}_k) \, \chi_k^*(\boldsymbol{\theta}_k)\right] \ , \tag{3.6}$$

where $\boldsymbol{\theta}_k$ and $\chi_k$ are the position and the ellipticity of the $k$-th galaxy image, $1 \leq k \leq N_{\mathrm{gal}}$, $n$ is the surface number density of the galaxies, and $W$ is the window function

$$W(\mathbf{x}; s) = 1 - \left(1 + \frac{x^2}{2s^2}\right) \exp\left(-\frac{x^2}{2s^2}\right) \ . \tag{3.7}$$

The window function is introduced to reduce the statistical noise at the point of reconstruction arising from the random positions of the galaxies and from the random intrinsic source ellipticities: From Eqs.(3.6) & (2.10) we see that the contribution from each galaxy $k$ to $\kappa(\boldsymbol{\theta})$ is inversely proportional to its squared distance $(\boldsymbol{\theta} - \boldsymbol{\theta}_k)^2$ to $\boldsymbol{\theta}$. As was shown in KS, if the weight function $W$ is left out of Eq. (3.6), the resulting estimate for $\kappa$ has infinite noise, due to the discreteness of galaxy images and the singular behavior of the kernel $\mathcal{D}(\boldsymbol{\theta})$ for $\theta \to 0$. The weight function smoothes over this singularity and yields an estimate for $\kappa$ with finite noise. We shall call Eq.(3.6) the 'KS-inversion formula', or the KS method.

In Fig. 3 we show the reconstructed surface mass density of the model cluster (Fig. 1) obtained with Eq.(3.6). The best result is expected if there are only circular sources, since then $\chi_k(\boldsymbol{\theta}_k) = \langle\chi\rangle(\boldsymbol{\theta}_k)$ and one source of noise is eliminated. Hence, we randomly distribute 3360 circular sources on an area that corresponds to $(10')^2$ on the lens plane and use a window function with $s = 0.'2$. We find that the reconstruction fails in the cluster center, where the model cluster has high surface mass densities. The reason for this is that the KS method can not be applied to the nonlinear lensing regime, since the underlying assumption (3.1) is no longer satisfied. We find this failure for all reconstructions in this section and we do not further comment on it. Next, we find that the mass increases towards the edges, and that even negative mass densities occur. This is due to the restriction of the integration (or summation) area from $\mathbb{R}^2$ to a finite region. In Sect. 4 we describe one method to eliminate these boundary effects, and again we will not comment further on it in this section. Finally, we find that for $\kappa \lesssim 0.5$ the reconstruction is relatively good besides small fluctuations which arises from the noise introduced from the random galaxy positions ('shot noise').

To reduce this noise, we introduce a regular quadratic grid with $41 \times 41$ points $\boldsymbol{\theta}_{ij}$ and calculate the mean ellipticity of neighbouring galaxies at each gridpoint

$$\bar{\chi}(\boldsymbol{\theta}_{ij}) = \frac{\sum_{k=1}^{N_g} w_k \, \chi(\boldsymbol{\theta}_k)}{\sum w_k} \ , \qquad \mathrm{with} \quad w_k = \exp\left(-\frac{(\boldsymbol{\theta}_{ij} - \boldsymbol{\theta}_k)^2}{(\Delta\theta)^2}\right) \ , \tag{3.8}$$

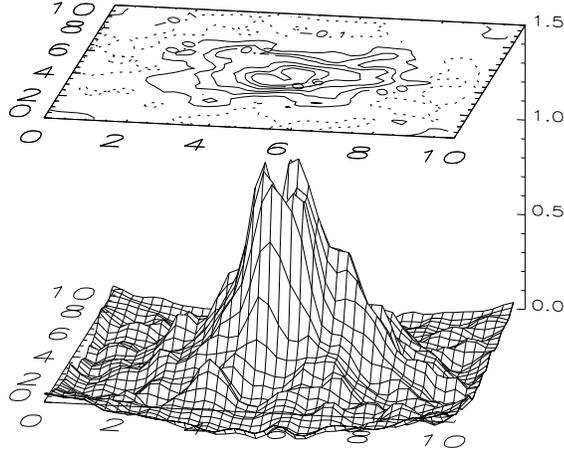

**Fig. 3.** The reconstructed surface mass density of the model cluster (Fig. 1) obtained with the KS inversion formula (Eq. (3.6)) with $s = 0\rlap{.}'2$ in Eq. (3.7). To demonstrate that the noise is caused by the random positions of the distorted galaxy images we here assume circular source galaxies for the inversion to eliminate the noise contribution from the source ellipticities

where $\Delta\theta$ is the smoothing length. We point out that this mean $\bar{\chi}$ over galaxies is an estimate for the expectation value $\langle\chi\rangle$ used in Eq.(3.5). Hence, the reconstruction of $\kappa(\boldsymbol{\theta})$ is performed by summing over the gridpoints

$$\kappa(\boldsymbol{\theta}) \approx \frac{a^2}{2\pi} \sum_{i,j} \mathcal{R}\mathrm{e}\left[\mathcal{D}(\boldsymbol{\theta} - \boldsymbol{\theta}_{ij})\ \bar{\chi}^*(\boldsymbol{\theta}_{ij})\right] \quad, \tag{3.9}$$

where $a$ is the separation of the grid points. Fig. 4 shows the result obtained with Eq.(3.9). We use the same circular source galaxies with the same image positions as in Fig. 3 and a smoothing length of $\Delta\theta = 0\rlap{.}'2$. Comparing Figs. 3 & 4 with Fig. 1, we see that in Fig. 4 smaller mass peaks can be identified. Hence, the difference between the KS estimate (3.6) and our estimate (3.9) is that KS introduced smoothing through a window function in the kernel, whereas we perform the smoothing in the distortion distribution. In the latter case, the shot noise is greatly reduced.

Next we consider non-circular sources with an ellipticity distribution according to Eq.(2.22), with $R = 0.3$. Again, we have the same image positions and we perform the reconstruction with Eq. (3.9), with $\Delta\theta = 0\rlap{.}'2$. As shown in Fig. 5, the reconstruction is noisy again, but the noise now comes in because $\bar{\chi}(\boldsymbol{\theta}_{ij})$ fluctuates around $\langle\chi\rangle$ which enters Eq. (3.5). However, we can reduce the noise using an adaptive smoothing length in Eq. (3.8), e.g.,

$$\Delta\theta_{ij} \sim 0.3 \times \left(1.05 - |\bar{\chi}_{i,j-1}|\right)^{0.75} \quad, \tag{3.10}$$

where $\bar{\chi}_{i,j-1}$ is the mean ellipticity found for the previous point of reconstruction; we start with $\Delta\theta = 0\rlap{.}'3$ at the boundaries. The result is shown in Fig. 6 and a comparison with Fig. 4 shows that now the reconstruction has almost the same quality as that for circular sources. The adopted smoothing length has approximately the effect of an adopted size

$$s(\boldsymbol{\theta}) \sim \frac{0.1}{\sqrt{\kappa^2 + 0.4^2}} \tag{3.11}$$

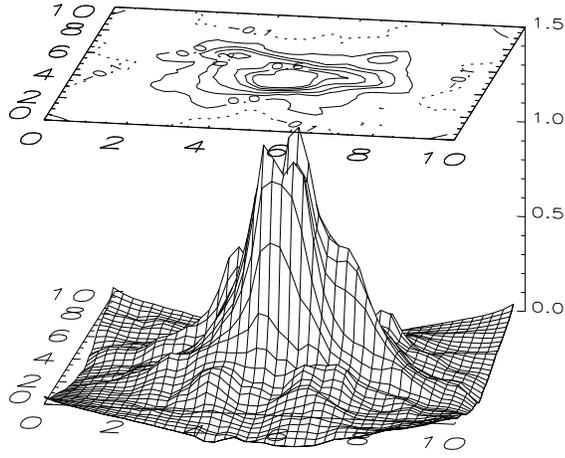

**Fig. 4.** The reconstructed surface mass density obtained with Eq.(3.9) using the same circular galaxies as for the reconstruction shown in Fig. 4. Calculating $\bar{\chi}(\boldsymbol{\theta}_{ij})$ according to Eq.(3.8) with $\Delta\theta = 0\rlap{.}'2$ and summing over regularly spaced gridpoints in Eq. (3.9) instead of summing over individual galaxies as done in Eq. (3.6) clearly removes the noise caused from the galaxy positions

for the window function $W$ in the KS inversion formula (3.6). But in the latter case the shot noise cannot be removed without simultaneously smoothing away small mass structures (in our example, we would remove the smallest of the three substructures).

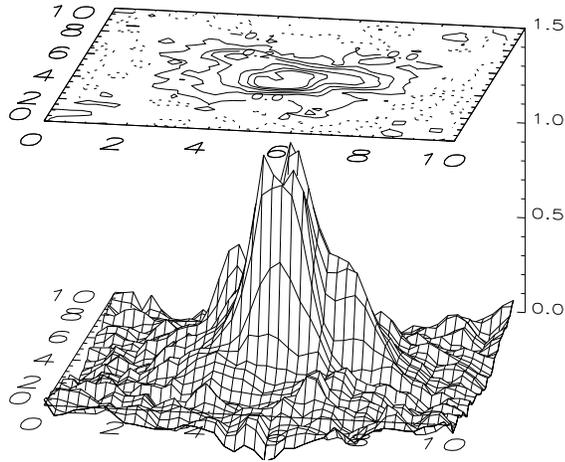

**Fig. 5.** The surface mass density is reconstructed in the same way as the surface mass density shown in the previous Fig. 4. We use the same image positions but non-circular sources with a ellipticity distribution according to Eq.(2.22) with $R = 0.3$. The smoothing length in (3.8) is chosen again to be $\Delta\theta = 0\rlap{.}'2$. In regions of small surface mass density (or better, of small distortions) this ellipticity distribution introduces noise, since $\bar{\chi}(\boldsymbol{\theta}_{ij})$ defined in Eq.(3.8) strongly fluctuates around the expectation value $\langle\chi\rangle(\boldsymbol{\theta}_{ij})$

As a final remark to inversion techniques in the weak lensing regime, we point out that the approximation (3.4) can be dropped if one determines $\delta(\boldsymbol{\theta}_{ij})$ from (2.21) instead of using the approximation (3.4), and applying the inversion formula (3.2) instead of (3.9). This approach is only slightly more complicated. In this case, we replace the integral in (3.2) by the sum over gridpoints and obtain

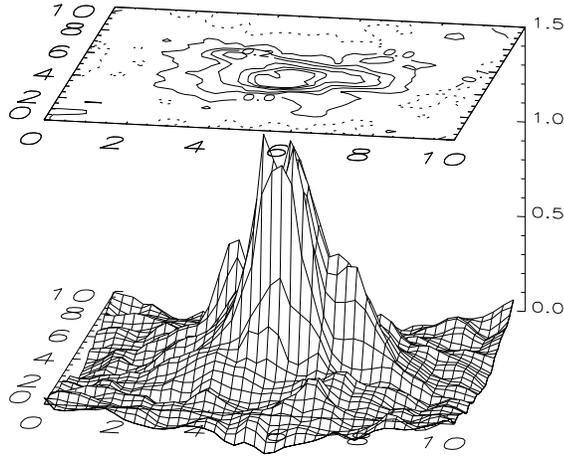

**Fig. 6.** The surface mass density obtained with Eq. (3.9), using the same source galaxies as for the reconstruction in Fig. 5 ($R = 0.3$) and the adaptive smoothing length of Eq.(3.10) for calculating $\bar{\chi}(\boldsymbol{\theta}_{ij})$ in Eq.(3.8)

$$\kappa(\boldsymbol{\theta}) \approx \frac{-a^2}{2\pi} \sum_{i,j} \mathcal{R}e\left[\mathcal{D}(\boldsymbol{\theta} - \boldsymbol{\theta}_{ij})\ \delta^*(\boldsymbol{\theta}_{ij})\right] \quad . \tag{3.12}$$

The resulting surface mass density deviates only slightly from that in Fig. 6 and is not shown here. The basic reason why there are no larger changes in the resulting $\kappa$ is that for the chosen ellipticity distribution of the sources the function $\zeta$ is very close to unity for all values of $\delta$. If it turns out that the true ellipticity distribution of faint background galaxies is shifted towards higher ellipticities, the difference in the two approaches can become significant.

To summarize, Fig. 6 is, besides the boundary effects, the best reconstruction that can be done for this cluster within the weak lensing approximation (3.1). If we compare the original mass distribution shown in Fig. 1 with the reconstructions shown in Fig. 6 and Fig. 3, where we simplified the reconstruction because of using only circular sources, then we find that the reconstruction shown in Fig. 6 is obviously better than that in Fig 3.

## 4. A simultanous reconstruction in the linear and nonlinear lensing regime

### 4.1 The inversion method

All Figs. 3 through 6 demonstrate that the central region with high surface mass density is badly reconstructed. This is due the approximation (3.1), which is obviously not valid in the central region of the cluster. Therefore, we drop this approximation and use Eq.(2.14) for the reconstruction. Again, we replace the integral over the $\mathbb{R}^2$ with a summation over a grid that covers the data field and obtain

$$\kappa(\boldsymbol{\theta}) = \frac{-a^2}{\pi} \sum_{i,j} \mathcal{R}e\left[\mathcal{D}(\boldsymbol{\theta} - \boldsymbol{\theta}_{ij})\, \gamma^*(\boldsymbol{\theta}_{ij})\right] \quad , \tag{4.1}$$

where $a$ is the gridspacing of the quadratic grid,

$$\gamma(\boldsymbol{\theta}_{ij}) = \frac{\left(1 - \mathrm{sg}(\boldsymbol{\theta}_{ij}) \sqrt{1 - |\delta(\boldsymbol{\theta}_{ij})|^2}\right)}{\delta^*(\boldsymbol{\theta}_{ij})} (1 - \kappa(\boldsymbol{\theta}_{ij})) \quad , \tag{4.2}$$

is the shear at the gridpoints, and $\mathrm{sg}(\boldsymbol{\theta}_{ij}) = \mathrm{sign}\left[\det A(\boldsymbol{\theta}_{ij})\right] = \pm 1$ is the (a priori unknown) sign of the Jacobian determinant at $\boldsymbol{\theta}_{ij}$. Eq.(4.1) can be solved iteratively: we start with $\kappa^{(0)}(\boldsymbol{\theta}_{ij}) = 0$ and $\mathrm{sg}^{(0)}(\boldsymbol{\theta}_{ij}) = 1$ for all grid points; then we calculate for $n \geq 1$

$$\kappa^{(n)}(\boldsymbol{\theta}) = \frac{-a^2}{\pi} \sum_{i,j} \mathcal{R}\mathrm{e}\left[\mathcal{D}(\boldsymbol{\theta} - \boldsymbol{\theta}_{ij})\, \gamma^{*(n-1)}(\boldsymbol{\theta}_{ij})\right] \quad , \tag{4.3}$$

with

$$\gamma^{(n)}(\boldsymbol{\theta}) = \frac{-a^2}{\pi} \sum_{i,j} \mathcal{D}(\boldsymbol{\theta} - \boldsymbol{\theta}_{ij})\, \kappa^{(n)}(\boldsymbol{\theta}_{ij}) \quad . \tag{4.4}$$

The function $\mathrm{sg}^{(n)}$ for $n \geq 1$ is obtained from the shear $\gamma^{(n)}$ and the surface mass density $\kappa^{(n)}$ as

$$\mathrm{sg}^{(n)}(\boldsymbol{\theta}_{ij}) = \mathrm{sign}\left[\det A^{(n)}(\boldsymbol{\theta}_{ij})\right] = \mathrm{sign}\left[(1 - \kappa^{(n)}(\boldsymbol{\theta}_{ij}))^2 - \left|\gamma^{(n)}(\boldsymbol{\theta}_{ij})\right|^2\right] \quad . \tag{4.5}$$

For a non-critical cluster this iteration works straightforwardly and $\kappa^{(n)}$ converges quickly to the 'correct' value of the surface mass density $\kappa$. But for a critical cluster the kernel of (4.1) has a (formal) singularity for $|\delta(\boldsymbol{\theta}_{ij})| \to 0$ and $\mathrm{sg}(\boldsymbol{\theta}_{ij}) = -1$. Because of this, the numerical iteration becomes highly unstable and does not converge. We found out that this instability can be removed if we multiply each term $ij$ in the sum of Eq.(4.3) with $G^{(n-1)}(\boldsymbol{\theta}, \boldsymbol{\theta}_{ij})$, where

$$G^{(n)}(\boldsymbol{\theta}, \boldsymbol{\theta}_{ij}) = W\left(\boldsymbol{\theta} - \boldsymbol{\theta}_{ij}, \frac{a}{2}\right) \left(1 + \frac{1}{2}\left|\gamma^{(n)}(\boldsymbol{\theta}_{ij})\right|^2\right) \exp\left(-\frac{1}{2}\left|\gamma^{(n)}(\boldsymbol{\theta}_{ij})\right|^2\right) \quad , \tag{4.6}$$

where $W(\boldsymbol{\theta}, s)$ is given by (3.7). Then Eq.(4.3) reads

$$\kappa^{(n)}(\boldsymbol{\theta}) = \frac{-a^2}{\pi} \sum_{j,i} \mathcal{R}\mathrm{e}\left[\mathcal{D}(\boldsymbol{\theta} - \boldsymbol{\theta}_{ij})\, G^{(n-1)}(\boldsymbol{\theta}, \boldsymbol{\theta}_{ij})\, \gamma^{*(n-1)}(\boldsymbol{\theta}_{ij})\right] \quad . \tag{4.7}$$

The reason for the choice of the function $G$ is the following: for $\delta \to 0$, Eq.(2.13) yields $\gamma \to (1 - \kappa)\delta/2$ for subcritical ($\mathrm{sg} = +1$) and $\gamma \to 2(1 - \kappa)/\delta^*$ for supercritical regions ($\mathrm{sg} = -1$) in the lens plane; since in the latter case $\kappa \to 1$, the shear $\gamma$ remains finite for $\delta \to 0$. In our iterative reconstruction, the shear $\gamma$ is approximated by $\gamma^{(n)}$ at each reconstruction step. But since we will not necessarily have $\kappa^{(n)} \to 1$ for $\delta \to 0$ and $\mathrm{sg} = -1$, we may have rather high values for $\gamma^{(n)}$ at the beginning of the reconstruction. We cut off this high values for $\gamma^{(n)}$ by multiplying $\gamma^{(n)}$ with $(1 + \frac{1}{2}\left|\gamma^{(n)}\right|^2) \exp(-\frac{1}{2}\left|\gamma^{(n)}\right|^2)$. This motivates the second part of the function $G^{(n)}$ in Eq.(4.6), and the reason for the first is the following: the grid causes artifical modes in Fourier space, and we supress all modes with $k > 2\pi/a$ using the transfer function (see KS Sect. 2.2) $T(k) = \exp(-k^2 a^2/8.)$. This corresponds to a window function $W$ with size $s = a/2$. If we replace $a/2$ by $a$ in the window function $W$, the smaller peaks of the density profile in the resulting mass reconstruction would be smoothed out, whereas if we used $a/4$, artifical local peaks can occur. The number of iterations necessary to achieve 'convergence' is about twice the number $N$ of gridpoints per dimension; each iteration step is performed very quickly, so that the numerical effort for the inversion is negligible.

To quantify the quality of the reconstruction we calculate

$$\chi^2 = \sum_{i=1}^{N_{\rm gal}} \left( \frac{\chi_i - \langle \chi \rangle (\boldsymbol{\theta}_i)}{\sigma_\chi} \right)^2 = \sum_{i=1}^{N_{\rm gal}} \left( \frac{\chi_i + \zeta(\delta(\boldsymbol{\theta}_i))\delta(\boldsymbol{\theta}_i)}{\sigma_\chi(\delta(\boldsymbol{\theta}_i))} \right)^2 \; , \qquad (4.8)$$

with

$$\sigma_\chi \approx \sqrt{M_2}\left(1-\delta^2\right)^{\mu_2} \qquad (4.9)$$

$$\zeta(\delta) \approx 1 - \frac{M_2}{2}(1-\delta^2)^{\mu_1} \quad ,$$

(Paper I Eqs. (3.22&23)) and

$$M_2 = R^2 - \frac{1}{e^{1/R^2}-1} \; , \quad M_4 = 2R^2 - \frac{1+2R^2}{e^{1/R^2}-1} \; ,$$

$$M_6 = 6R^6 \frac{1+3R^2+6R^4}{e^{1/R^2}-1} \; , \quad \mu_1 = 1 - \frac{3M_4}{4M_2} \; , \quad \mu_2 = \frac{6M_2 + M_2^2 - 9M_4}{8M_2} \; ,$$

from the reconstructed mass density $\kappa^{(n)}$ for each iteration step and show the results for three different realizations of source distributions in Fig. 7 (for our choice of $R$ in (2.22), we get $\mu_1 \approx 0.86$ and $\mu_2 \approx 0.56$). For some realizations of the source distributions the iteration algorithm (4.7) finds a stable solution (solid curve in Fig. 7) or it finally iterates between very few similar mass profiles and the $\chi^2$ becomes periodic as shown by the dotted curve in Fig. 7. For this simulation the maximum change of the surface mass density at a gridpoint was $\Delta \kappa_{\max} \lesssim 0.03$. These profiles $\kappa$ have different $\text{sg}^{(n)}(\boldsymbol{\theta}_{ij})$ for one or very few gridpoints $\boldsymbol{\theta}_{ij}$ near the center. As we already pointed out, one cannot determine the sign locally due to the local degeneracy (2.13). Therefore, one can obtain more than one solution for $\kappa$. For other iterations we find that the $\chi^2$ fluctuates strongly even after more than 100 iteration steps. One reason for this is that one single galaxy close to a critical curve (where $|\delta| = 1 - \epsilon$, $\epsilon \to 0$) can strongly change $\chi^2$ if there is a small change in $\delta$ (i.e. a small change in the surface mass density) since $\sigma_\chi \propto \epsilon^{\mu_2}$ and $\chi_i - \langle \chi \rangle (\boldsymbol{\theta}) \approx c-\epsilon$, where $c$ is a small constant. Another reason is that if sign$(\boldsymbol{\theta}_{ij})$ changes at one or more gridpoints than the algorithm needs a few steps of iterations to adjust the surface mass density to the new critical curve, therefore $\chi^2$ is increased until the surface mass density fits to the critical curve again. If the sign changes frequently for several gridpoints then strongly fluctuating values for $\chi^2$ occur. An example for this is shown with the dashed curve in Fig. 7. In this case we choose the surface mass density which gives the smallest value for $\chi^2$. Fig. 7 shows that the values for $\chi^2$ are clearly above the expectation value for a perfect reconstruction, i.e. $\chi^2 \approx N_{\rm gal} = 3360$. The relatively high value of $\chi^2$ is dominated by a few galaxy images close to the critical curves. Due to the fact that our method includes smoothing and (bilinear) interpolation between grid points for the calculation of $\delta$ at the galaxy positions, we cannot expect to get accurate values of $\delta$, and since $\sigma_\chi$ becomes very small for $1 - |\delta| \ll 1$, the corresponding galaxy images lead to a fairly large value of the formal error. We expect that by further modifications of our method (such as a further 'dynamical' adoption of the smoothing length to the reconstructed $\delta$, or a grid with varying spacing) the value of $\chi^2$ can be reduced; we hope to return to such refinements in a later paper.

All the above mentioned problems do not occur if we consider a non-critical cluster. There, no local degeneracy exists, and since there are no critical curves, the value of $\sigma_\chi$ can never become very small. In fact, we have reconstructed the same cluster as shown in Fig. 1, but with a surface mass density scaled down by a factor of two. In this case, our iteration process converges after a small number ($\leq 5$) of iteration steps, and the corresponding value of $\chi^2$ is significantly smaller than the number of galaxy images. Hence we conclude that our iterative method works without any problem for noncritical clusters, and that the above mentioned problems are solely caused by the critical curves and the corresponding local degeneracy of the sign in (4.2).

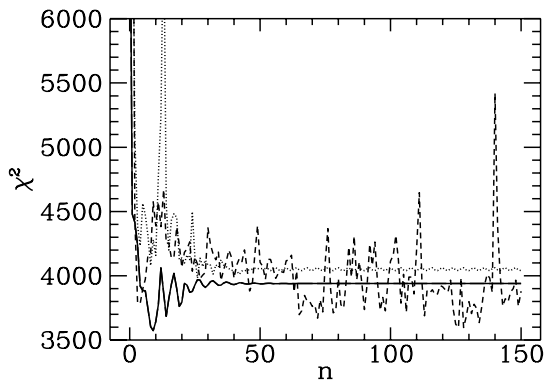

**Fig. 7.** The $\chi^2$ value (Eq. (4.8)) obtained for three different realizations of the source distribution as a function of the iteration step n. For explanation see text

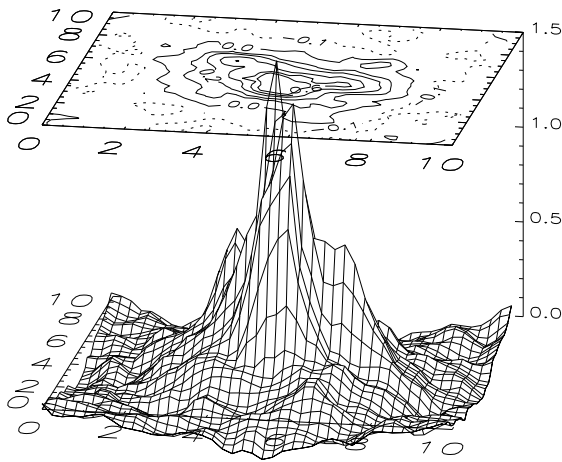

**Fig. 8.** The reconstructed surface mass density obtained with Eq.(4.7)

In Fig. 8 we show the result of this reconstruction method for the same realization of the source distribution as used in Figs. 3 to 6. In general, the height of the mass peaks is still smaller than that of the input mass distribution (see Fig. 1) which is expected since smoothing reduces the height of peaks, but the reconstruction is highly improved compared to Fig. 6. In fact, given the difficulties we had in reconstructing a spherical cluster in Paper I, it is surprising that the iteration procedure finds a solution which is very similar to the original mass profile. The reason for this is probably the introduction of the factor $G$ in (4.7), which guarantees a smooth iteration. As we pointed out in Paper I, smoothness is the only way to distinguish between the many solutions (which differ only by the sign in (4.2)) possible for a given distortion distribution.

Our simulations demonstrate that distorted background images (not necessarily arcs, but also arclets) can be used to obtain the surface mass density distribution and the critical lines of supercritical clusters. However we stress that there is a global invariance transformation (Paper I, Sect. 3.4): if $\kappa_0(\boldsymbol{\theta})$ is a reconstruction of the mass distribution which fits the deduced distortions, then

$$\kappa(\boldsymbol{\theta};\lambda) := 1 - \lambda + \lambda\kappa_0(\boldsymbol{\theta}) \;, \tag{4.10}$$

is an equally good reconstruction for any $\lambda$. Hence, a solution $\kappa_0$ is not unique, as long as one has no further information about the mass distribution (e.g., power law decline for large $|\boldsymbol{\theta}|$, or the method proposed by Broadhurst, Taylor & Peacock 1994, which is based on the magnification of the background galaxies).

To see whether the reconstructed mass distribution matches the distortion $\delta$ deduced from the observations, we calculate $\gamma$, according to (4.4), from the reconstructed surface mass density shown in Fig. 7 and then $\delta$ according to Eq. (2.12), and show $\delta_2 = \mathcal{I}\mathrm{m}(\delta)$ in Fig. 9, which can be compared with Fig. 2. The noise in Fig. 9 is significantly reduced relative to that of Fig. 2b, and is in good agreement with the original distortion shown in Fig. 2a, but close to the boundaries the slope of the decline of $|\delta_2|$ is a bit too steep compared to Fig. 2a,b. The basic reason for this is the finite size of the data field, which enters twice in the reconstruction of $\delta$: first, in the calculation of $\kappa$ from the shear distribution $\gamma$, and second, in the calculation of the shear $\gamma$ from $\kappa$, according to Eqs. (4.3 & 4). In other words, whereas the pair of equations (2.9 & 11) are exact inversions of each other, this is no longer true if the integration, or in our case the summation, is performed over a finite area. This then leads to the boundary effect which can be seen in Fig. 7 and which is discussed next in greater detail.

### 4.2 Boundary effects

To demonstrate the influence of the geometry of the data field (i.e., the shape of the CCD) on the reconstructed mass density, we choose a rectangular data field with sidelengths $6' \times 10'$, which has a shape similar to that used by Smail et al. (1994b) for determining the mass distribution in the cluster 0016+16 ($z_d = 0.55$). They use the KS inversion formula Eq.(3.6) with a constant smoothing length $s = 0'.45$ in the window function Eq. (3.7) and show their results in their Figs. 7 & 8a.

We apply the same steps on our cluster and show the results in Fig. 10 for $s = 0'.45$ (a) and $s = 0'.2$ (b). On the one hand small structures are smoothed out in Fig. 10a because a smoothing length of $s = 0'.45$ is too large, and on the other hand regions with high mass densities are badly reconstructed, as expected, because the weak lensing approximation used for the KS method is not valid here. But what is more important here is that Fig. 10a shows two additional mass structures at the top and at the bottom.

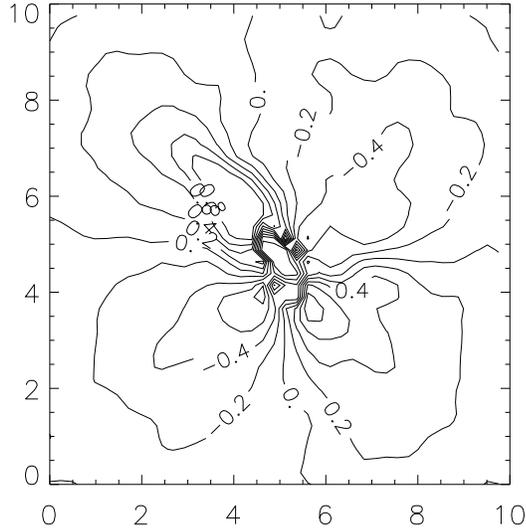

**Fig. 9.** The distortion-component $\delta_2 = \mathcal{I}m(\delta)$ deduced from the reconstructed surface mass density shown in Fig. 7 using Eqs.(4.3) & (2.12)

These are obviously artefacts arising from the shape of the choosen data field. These artifacts can also been seen in the corresponding figures in Smail et al. (1994b). To keep these effects as small as possible or even to avoid them at all we suggest the following:
(1) In the absence of any correction for the shape and size of the data field, one should try to use a circular or quadratic field. The larger the field, the smaller the boundary effects will be.
(2) Extrapolate the distortion field to the outside of the observation field, such that the remaining boundary effects are neglegible. We will illustrate this method below.
(3) One can modify the inversion equation in such a way that the finite size and shape of the data field is accounted for, so that the resulting inversion equation is exact on a finite field; we are currently working on finding such an inversion equation.

Here we demonstrate the principle of how method (2) works in the case of our cluster. From Fig. 7 we realize that the mass distribution in the outer parts of the cluster can be roughly approximated by an axially-symmetric mass distribution. Because of this we use all gridpoints with separation larger than 2′ from the central mass peak, and fit a mass distribution with

$$\kappa^{\mathrm{model}}(\theta) = \frac{\kappa_0}{(\theta^2 + \theta_c^2)^{p/2}} \tag{4.11}$$

to the distortion $\delta$ found at those gridpoints. We determine the parameters $\kappa_0, \theta_c, p$ which minimize $\sum (\delta^{\mathrm{model}} - \delta)^2$ and use these to extrapolate the distortion to the outside of the data field. The total sidelength of the combined extrapolated and observed distortion field is twice that of the original data field. Fig. 11a shows a contourplot for the imaginary part $\delta_2 = \mathcal{I}m(\delta)$ of the combined distortion field.

Next we apply Eq.(4.7) on this enlarged distortion field, solve it again iteratively and show the result in Fig. 12. Comparing Figs. 7 & 12 we find that an enhanced distortion field largely removes the boundary artefacts. From Fig. 12 we recalculate the distortion $\delta$ and show $\delta_2$ in Fig. 11b for comparision with Fig. 11a. The fit to the extrapolated data

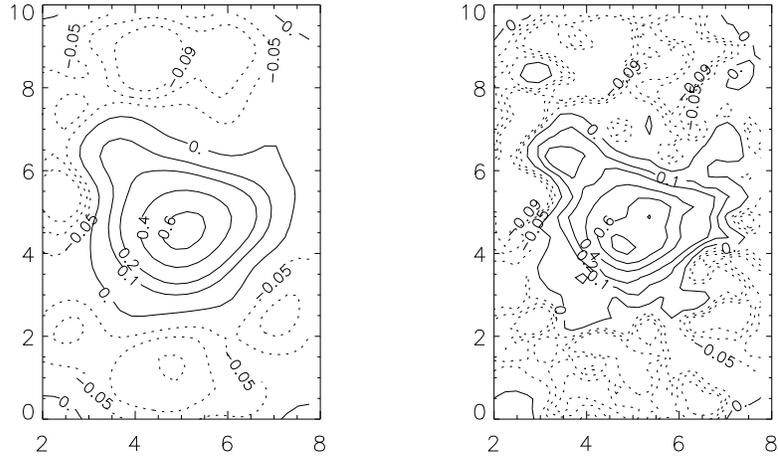

**Fig. 10.** A contour plot of the reconstructed surface mass density obtained with a rectangular data field of about $6' \times 10'$. Similar to Smail et al. (1994b) we use the KS inversion formula (3.6) with a constant smoothing scale of (a) $s = 0\rlap{.}'45$ (left panel) and (b) $s = 0\rlap{.}'2$ (right panel) for the window function (3.7). These figures should be compared with Figs. 7&8a of Smail et al. (1994b) and with the contourlines of our Fig. 2 where we used a quadratic data field and $s = 0\rlap{.}'2$. The two features at the ends of the long side of the rectangle are just artefacts of the shape of the data field and correspond to negative surface mass density (dotted contour lines). From the agreement of the shape of the countours with those of Smail et al. (1994b), we suspect that their reconstruction for 0016+16 also has an extended region of negative surface mass density

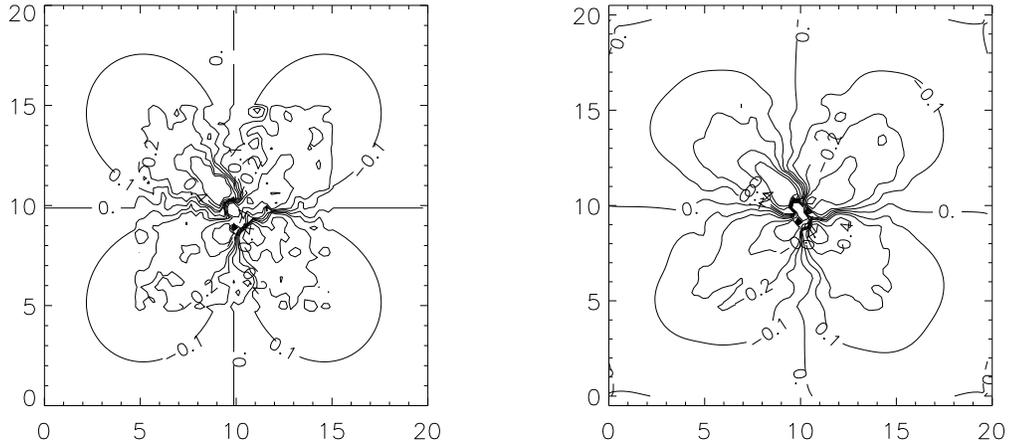

**Fig. 11.** (a)–left panel: The reconstructed distortion field $\delta_2 = \mathcal{I}\mathrm{m}(\delta)$ (see Fig. 2b) is extrapolated to the outside of the data field. For this we assume that for a distance $> 2'$ from the central mass peak, the mass distribution can be roughly described with Eq. (4.11). Then we use the reconstructed distortions outside of this area and estimate the parameters $\kappa_0, \theta_c, p$ – see (4.5) – by minimization of $\sum (\delta^{\mathrm{modell}} - \delta)^2$. Last, we use these parameters to calculate the distortion up to a distance that is twice the length of the data field. (b)–right panel: The imaginary part of the distortion field $\delta_2 = \mathcal{I}\mathrm{m}(\delta)$ obtained from the reconstructed mass distribution shown in Fig. 12

field is not perfect, since for example the shape of the contourlines with $|\delta_2| = 0.1$ is obviously different in Fig. 11a&b. The basic reason for this is that the extrapolation of the distortion $\delta$ is done with a fairly crude model, Eq. (4.11). In principle one could try to improve the reconstruction at the boundaries by using more complicated models for the extrapolation of $\delta$. But we do not want to investigate this method in more detail, since we hope to make progress soon in developing method (3) mentioned above, which then will be superior to the method investigated here.

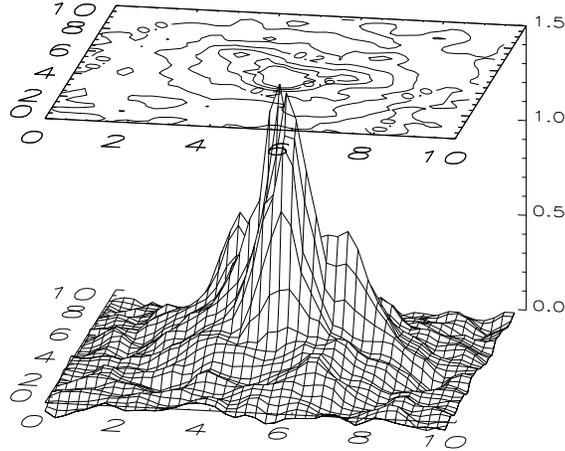

**Fig. 12.** The reconstructed surface mass density obtained with Eq.(4.7) using the enhanced distortion field shown in Fig. 11a

To quantify the quality of the reconstruction we calculate $\chi^2_{\rm org}$ (4.8) for the original mass density and compare it with $\chi^2_{\rm rek}$ for the reconstructed mass distribution for different realizations of the source distribution. The open squares in Fig. 13 show the results for those reconstructions where we have extrapolated the distortion field as described above, whereas the crosses correspond to the original data field only. We find that the extrapolation decreases $\chi^2_{\rm rek}$, but $\chi^2_{\rm rek}$ is still up to 15% higher than $\chi^2_{\rm org}$.

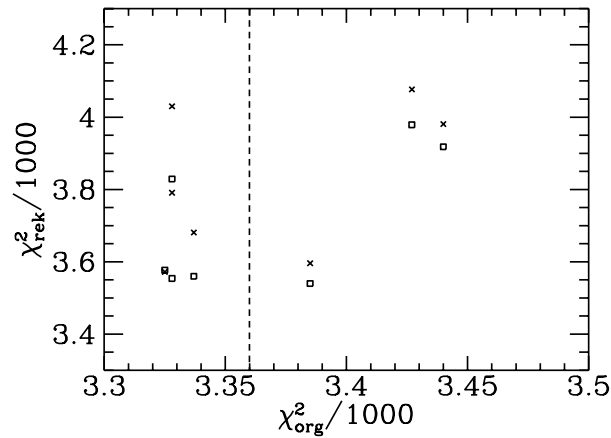

**Fig. 13.** The values for $\chi^2_{\rm rek}$ (4.8) of the reconstructed surface mass distribution (open squares: with, crosses: without extrapolation of $\delta$; see text!) as a function of $\chi^2_{\rm org}$ of the original mass distribution for several different realizations of the source population. The total number of source galaxies was fixed at $N_{\rm gal} = 3360$ but different random positions and random ellipticities were used. The dashed line indicates the expectation number of $\left\langle \chi^2_{\rm orig} \right\rangle = N_{\rm gal}$

### 4.3 Mass estimates

Comparing Figs. 1 & 12 we find that the mass distribution is reconstructed fairly accurately. To quantify the accuracy we now compare the total mass $m(\theta < \theta_g)$ within a distance $\theta_g$ from the cluster center found for the original and reconstructed mass densities. For both reconstructions shown in Figs. 7 & 12 we have negative values for $\kappa$ in cluster regions with small original mass densities. First, the finite data field leads to extended regions with negative mass densities (see blobs at the boundaries in Figs. 7 & 10), second, the noise in the reconstructed distortion $\delta$ (see Fig. 2b) leads to a noisy reconstruction of $\kappa$ in regions of small mass densities. To formally remove the unphysical negative mass densities, we use the global invariance transformation (4.10) and choose $\lambda$ such that the minimum value of $\kappa$ on our field is zero. For the mass distribution shown in Fig. 7 taken as $\kappa_0$ in (4.10), where we have not corrected for boundary artefacts, we obtain $\lambda \approx 0.17$, and for the mass distribution in Fig. 12 we obtain $\lambda \approx 0.08$. We show the resulting mass density obtained by applying (4.10) to the mass distributions of Fig. 12 in Fig. 14. Now the mass density of the outer parts of the cluster is overestimated and the central peak is too low, as can be seen by comparing with Fig. 1.

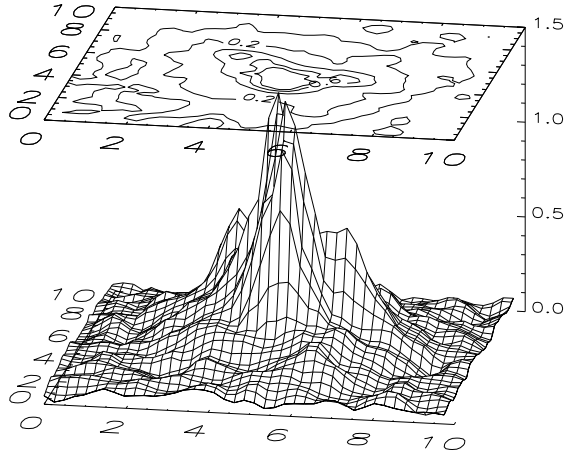

**Fig. 14.** The surface mass density obtained from Fig. 12 with the gloabal invariance transformation $\kappa = (1 - \lambda)\kappa_0 + \lambda$ with $\kappa_0$ taken from Fig. 12. $\lambda = 0.084$ is chosen such that the minimum of the resulting mass distribution $\kappa$ is zero

From the reconstructed surface mass densities and the above values of the scaling parameter $\lambda$, we calculate the mass $m(\theta < \theta_g)$ within a radius $\theta_g$ from the cluster center. The result in units of the original cluster mass within $\theta_g$ is shown in Fig. 15. The two solid curves show the result if we remove the negative mass densities with the invariance transformation, the two dashed lines the results if we directly use the mass distributions given in Figs. 7 & 12. Fig. 7 with $\lambda = 0$ gives the worst results because mass densities with rather negative values (down to $\kappa \approx -0.2$) occur. Fig. 12 with $\lambda = 0$ gives the best results, the mass is slightly underestimated in the center and decreases slightly too fast outwards, probably because the extrapolation of the distortion field is not sufficiently accurate (e.g., not far enough out).

The two solid lines in Fig. 15 are the only distributions with no unphysical negative mass densities and are very similar. We think that one could improve the upper most curve (i.e., decrease the corresponding value of $\lambda$) if one would use a smoothing function

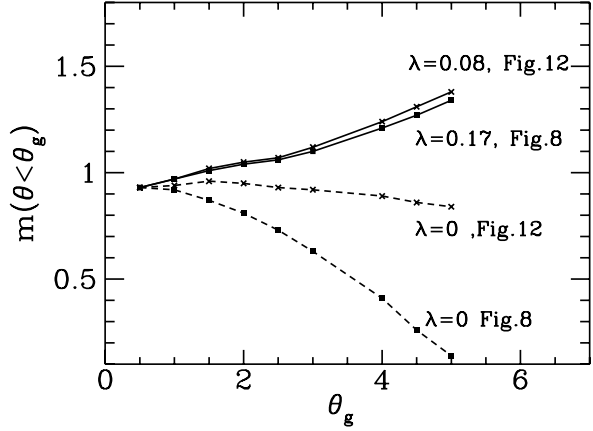

**Fig. 15.** The reconstructed mass $m(\theta < \theta_g)$ within a radius $\theta_g$ from the center of the cluster in units of the orginal mass within the same radius. The dashed curves show $m(\theta < \theta_g)$ for a direct mass determination from Figs. 7 & 12, the solid curves show $m(\theta < \theta_g)$ after applying the global invariance transformation (4.10) with the indicated values of $\lambda$

in (2.23) that reduces the amplitude of the noise for the reconstruction of small $\delta$. Then the reconstructed surface mass density in Fig. 12 would show less noise in regions of small $\kappa$, and because of this one would need a smaller value of $\lambda$ in the invariance transformation. (For example, if we could halven the noise then we would have $\lambda \approx 0.04$ and $m(\theta < 5') \approx 1.1$).

Given the large number of steps necessary to derive this final mass estimate, we are very pleased with its accuracy, in particular with that in the central region of the cluster. Although the application of this method to real data will most likely yield a higher error in the reconstruction of the density profile and, in particular, the total cluster mass, we are convinced that this method can yield far more reliable mass estimates than 'standard' methods, which are based on restrictive assumptions like symmetry, virial or hydrostatic equilibrium, etc.

## 5. Summary and discussion

In this paper we have developed a method to reconstruct the projected mass density of a cluster of galaxies using gravitationally distorted background galaxies, following the basic ideas of Kaiser & Squires (1993). We have modified their approach in several ways:
(1) We have applied the smoothing to the distortion data, instead of using a smoothed kernel in the inversion equation, as was done by KS. In this way, we have removed the shot noise (as demonstrated by comparing Figs. 3 & 4).
(2) The smoothing scale we used depends on the 'strength of the signal'; i.e., in regions where the distortion is strong, the ellipticity of the images is dominated by the lensing effect, so that the latter can be accurately obtained by averaging over a small number of images, whereas in regions of weak distortions, the noise introduced by the ellipticity of the sources requires a larger smoothing scale. In this way, we have a higher resolution in the inner parts of the cluster where the length scale of density variations are smaller, and lower resolution in the outer parts; this is illustrated by comparing Figs. 5 & 6.

(3) We have dropped the approximation $\delta \approx -\langle\chi\rangle$; this modification is important for source distributions with large mean ellipticities.
(4) We used the relation (2.13) between the shear and the distortion, the only observable, in the inversion eqation (2.14), instead of using the approximation $\gamma \approx \delta/2$, which is valid in the weak lensing regime ($\kappa \ll 1$, $|\gamma| \ll 1$) only. This modification is essential for reconstructing clusters with high surface mass densities, i.e. clusters which are critical, or nearly so, in their inner parts. The frequent occurrence of luminous arcs in clusters of galaxies shows that many X-ray luminous clusters (Hammer et al. 1994) are critical.
(5) We have obtained a fast iterative procedure to solve the inversion equation, which is an integral equation in general, and reduces to a simple integral only in the case of weak lensing. Examples of the results from this procedure can be seen in Figs. 12 & 14.
(6) We have pointed out the importance of, and made a first attempt to remove boundary effects, which occur due to the finite size of the data field (or CCD) – see Figs. 7 & 10.

Taken together, these modifications yield a substantial improvement of the original KS technique. The modifications (1-3) and (6) are already relevant for the weak lensing regime, whereas the other two points are essential for the strong lensing case. We have applied this method to a model cluster generated from a cosmological $N$-body simulation with a CDM-like spectrum. The resulting reconstruction of the mass distribution is surprisingly accurate when compared to the original mass distribution (compare Figs. 12&14 with Fig. 1!). In particular, the determination of the total mass within the inner few arcminutes of the cluster center is accurate at the few percent level, i.e., it is a surprisingly accurate estimate.

In this paper we have confined our consideration to the case that the angular diameter distance ratio $D_{\rm ds}/D_{\rm s}$ is almost the same for all sources. From the present knowledge of the redshift distribution of faint galaxies, we conclude that we can determine the projected mass distribution of a cluster with $z_d \lesssim 0.2$, if gravitationally distorted faint background galaxies can be observed. For higher redshift clusters one has to take into account the redshift distribution of the background galaxies. Whereas this distribution will probably somewhat complicate the analysis, it provides, on the other hand, a tool to study the redshift distribution of the faint galaxy population; a first step in this direction has been undertaken by Smail et al. 1994a.

*Acknowledgement.*We would like to thank Matthias Bartelmann and Matthias Steinmetz for providing us with the data from their cluster simulation.